\documentclass [aps,preprint,a4paper,floatfix]{revtex4}
\usepackage{amsfonts}
\usepackage{graphicx}
\usepackage{amsmath}
\usepackage{bm}
\usepackage{dcolumn}
\usepackage{hyperref}

\newcommand{\fm}{\text{fm}}
\newcommand{\MeV}{\text{MeV}}

\newcommand{\soge}{\text{SOGE}}

\begin{document}

\title{Dynamical study of the possible molecular state $X(3872)$ with
  the $s$-channel one gluon exchange interaction}

\author{Bao-Kai Wang} \affiliation{Department of Physics and State Key
  Laboratory of Nuclear Physics and Technology, Peking University,
  Beijing 100871, China}

\author{Xiao-Lin Chen} \affiliation{Department of Physics, Peking
  University, Beijing 100871, China}

\author{Wei-Zhen Deng} \email{dwz@th.phy.pku.edu.cn}
\affiliation{Department of Physics and State Key Laboratory of Nuclear
  Physics and Technology, Peking University, Beijing 100871, China}



\begin{abstract}
  The recently observed $X(3872)$ resonance, which is difficult to be
  assigned a conventional $c\bar{c}$ charmonium state in the quark
  model, may be interpreted as a molecular state. Such a molecular
  state is a hidden flavor four quark state because of its
  charmonium-like quantum numbers. The $s$-channel one gluon exchange
  is an interaction which only acts in the hidden flavor multi-quark
  system. In this paper, we will study the $X(3872)$ and other
  similiar hidden flavor molecular states in a quark model by taking
  into account of the $s$-channel one gluon exchange interaction.
\end{abstract}
\maketitle

\section{Introduction}

Recently, many new resonances are discoveried experimentally. Many of
them have the proper quantum numbers of the $q\bar{q}$ meson
states. However, their mass values do not fit the conventional
$q\bar{q}$ states in the quark model.  Among them, $X(3872)$ was first
observed in the $J/\psi\pi^+\pi^-$ channel by Belle collaboration in
2003 \cite{Choi:2003ue}, and has been confirmed by CDF
\cite{Acosta:2003zx}, D0 \cite{Abazov:2004kp} and Babar collaborations
\cite{Aubert:2004ns}.  Its quantum numbers are probabely $J^{PC}=1^{++}$. The
corresponding charmonium candidates in the quark model are $2^3P_1(3990)$ and
$3^3P_1(4290)$ which are $50\sim 200$ MeV above $M_X=3872$ MeV.

Many people suggested that $X(3872)$ is mainly a $D\bar{D}^*$
molecular state \cite{Close:2003sg, Voloshin:2003nt, Swanson:2003tb,
  Wong:2003xk, Tornqvist:2004qy}.  However, to bind the quarks and
anti-quarks together in such a four quark state or other multi-quark
states, we need to introduce new interaction into the quark model.
Swanson proposed that the $X(3872)$ is mainly a $D\bar{D}^*$ molecule
bound by the meson-meson interaction derived from the one pion
exchange and the quark exchange \cite{Swanson:2003tb}. In Wong's work
\cite{Wong:2003xk}, the meson-meson interaction is derived from a
QED-type effective interaction in terms of effective charges for
quarks and antiquarks.  In refs.~\cite{AlFiky:2005jd, Fleming:2007rp,
  Braaten:2007ct, Hanhart:2007yq, Voloshin:2007hh, Colangelo:2007ph},
further investigations based on the molecule assumption were carried
out.

Since the strength of the one pion exchange interaction seems not
strong enough to bind the $D\bar{D}^*$ molecular state, other authors
argued that $X(3872)$ may be a dominant $c\bar{c}$ charmonium with
some admixture of $D\bar{D}^*$ \cite{Suzuki:2005ha, Meng:2005er,
  Zhu:2007wz}. In ref.~\cite{Liu:2008fh}, after taking into account of
the sigma meson exchange potential, the interpretation of $X(3872)$ as
a loosely bound molecular state was further disfavored.

We should notice that the color structure of a multi-quark state is
much richer than that of a $q\bar{q}$ conventional meson state. Unlike
the conventional mesons or baryons, the $q\bar{q}$ and $qq$ pairs in a
multi-quark state can be in the color $8_c$ and $6_c$ representations
respectively. Some color interactions which have no effects in the
$q\bar{q}$ or $q^3$ colorless system, may have significant
contribution in a multi-quark system. So the complete interactions in
the quark model can be quite different after we take into account of
these multi-quark states.

The $s$-channel one gluon exchange interaction is an interaction
between quark and anti-quark of the same flavor which annihilate into a
virtual gluon.  It has no effect on the conventional $q\bar{q}$ mesons
but acts in the hidden flavor multi-quark system like the
charmonium-like moelcular states. In this work, we will investigate
the hidden flavor molecular states by considering the $s$-channel
gluon exchange interaction in the quark model. In the next section, we
will model the potential of the $s$-channel one gluon exchange
interaction starting from the non-relativistic reduction. Then the
$J^{PC}$ quantum numbers of the molecular states are selected by an
analysis of the spin dependence of the interaction strength. In
sec.~\ref{sect-3}, we will carry out the numerical calculation of
$X(3872)$ and some other charmonium-like states as the molecular
states. Also we will make prediction about the similar bottomium-like
molecular states. Finally, we will give a brief summary.

\section{The Potential of $S$ Channel One Gluon Exchange Interaction}

In our work, we will use the Bhaduri quark model which is a rather
simple non-relativistic quark potential model.  In the Bhaduri model,
the hamiltonian can be written as\cite{Bhaduri:1981pn}
\begin{equation}
  \label{eq:1}
  H=\sum_i (m_{i}+\frac{\bm{P}_i^2}{2m_i})
  -\frac{3}{4}\sum_{i<j}
  \left(\bm{F}_i\cdot\bm{F}_jV_{ij}^{C}
    +\bm{F}_i\cdot\bm{F}_j\bm{S}_i\cdot\bm{S}_jV_{ij}^{SS}\right).
\end{equation}
$m_i$ are constituent quark masses and
$\bm{F}_i^c=\frac12\bm{\lambda}_i^c \ (c=1,...,8)$ are the well-known
$SU_c(3)$ Gell-Mann matrices.  Apart from a constant, here the central
potential is the usual one gluon exchange coulomb potential plus the
linear confinment:
\begin{equation}
  V_{ij}^{C}=-\frac{\kappa}{r_{ij}}+\frac{r_{ij}}{a_{0}^{2}}-M_0 .
\end{equation}
$r_{ij}=|\bm{r}_i-\bm{r}_j|$ is the distance between quark $i$ and
$j$.  The color-magnetic interaction reads
\begin{equation}
  V_{ij}^{SS}=\frac{4\kappa}{m_im_j}\frac{1}{r_0^2r_{ij}}e^{-r_{ij}/r_0} ,
\end{equation}
where the $\delta$-interaction has been smeared smoothly with the
prescription
\begin{equation}
  \label{smear-ansatz}
  \delta^3(\bm{r}) \to \frac{1}{4\pi r_0^2r_{ij}}e^{-r_{ij}/r_0}.
\end{equation}
The model parameter values are
\begin{align*}
  \kappa&=102.67\MeV\fm,& a_0&=0.0326(\MeV^{-1}\fm)^{\frac12}, \\
  M_0&=913.5\MeV,& r_0&=0.4545\fm \\
  m_u&=m_d=337\MeV,& m_s&=600\MeV,\\
  m_c&=1870\MeV,& m_b&=5259\MeV .
\end{align*}

\begin{figure}
  \caption{\label{fig-1}%
    The $s$-channel one gluon exchange}
  \[
  \includegraphics{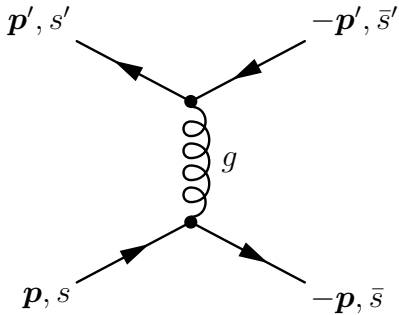}
  \]
\end{figure}

The $s$-channel one gluon exchange interaction (SOGE) takes place when a
$q\bar{q}$ pair annihilates into a virtual gluon
(Fig.~\ref{fig-1}). The non-relativistic reduction of the potential
is:
\begin{equation}
  V^{\soge} (\bm{r}_{ij}) = - \frac12 \left( \frac{4}{3} +\bm{F}_q
    \cdot \bm{F}_{\bar{q}} \right) \left( 1 + \frac{4}{3} \bm{S}_q
    \cdot \bm{S}_{\bar{q}} \right) G (4 m^2) \delta^3 (\bm{r}_{ij})
  \delta(f_i,f_j),
\end{equation}
where $G(4m^2)$ is the one gluon exchange amplitude.  The last
$\delta(f_i,f_j)$ represents that the quark and the anti-quark must
have the same flavor. Clearly, the two factors in the brackets mean
that this interaction only occurs when the $q\bar{q}$ pair is in the
color octet and with spin $S=1$. In our model calculation, the
$\delta$-interaction should be smeared smoothly with the same
prescription (\ref{smear-ansatz}) used in the color-magnetic
interaction in the same quark model. However, the QCD amplitude
$G(4m^2)$ is taken value in the timelike region, where the QCD
behavior is still not clear at present time.  In the spacelike region,
this amplitude is well known from the gluon propagator in perturbative
QCD and it reads
\begin{equation}
  G^t_{\text{pert}}(q^2) = \frac{4\pi\alpha_s}{-q^2} \qquad \text{for } q^2 <0 . 
\end{equation}
However, in our study of the hidden flavor molecular states, in order to
provide an attrative interaction to favor the formation of the bound
states, we need that
\[
G(4m^2) >0,
\]
which means the above formula from perturbative QCD should not be
directly used in the timelike region.  This change of sign is first
suggested in ref.~\cite{Li:1994ys} in the study of $\pi\pi$ and $K\pi$
scattering. Following ref.~\cite{Li:1994ys}, we assume that
\begin{equation}
G(4m^2) = -f G^t_{\text{pert}}(4m^2) = f \frac{\pi\alpha_s^2}{m^2},
\end{equation}
where $f$ is an introduced strength factor.  In our model,
after the $\delta$-function being smeared smoothly, the SOGE
potential turns to be:
\begin{equation}
  V_{ij}^{\soge}= - \frac{f}4 \left( 1 +\frac34 \bm{F}_q
    \cdot \bm{F}_{\bar{q}} \right) \left( \frac34 + \bm{S}_q
    \cdot \bm{S}_{\bar{q}} \right) 
  \frac{4\kappa}{m_i^2}\frac{1}{r_0^2r_{ij}}e^{-r_{ij}/r_0} \delta(f_i,f_j).
\end{equation}

First we observe that the potential $V^{\soge}$ is proportional to
$m_i^{-2}$, so the interaction mainly takes place between the light
$q\bar{q}$ pairs of $q=u,d,s$ in any multi-quark system. 

Next, we will analyse the spin dependence of $V^{\soge}$ in each
$J^{PC}$ channel of molecular state. For the molecular states we
concern, let us assume that their flavor structure is
$Q\bar{q}q\bar{Q}$, where $Q=c,b$ and $q=u,d,s$.  Since we can neglect
the $V^{\soge}$ interaction between $Q\bar{Q}$, the interaction acts only
when the $q\bar{q}$ pair is in color octet and $S=1$ as we have
mentioned before.  More specifically, the interaction will be switched
on if the spin coupling of the four quarks is
\[
[(Q\bar{Q})_{J_1} (q\bar{q})_{J_2=1}]_J.
\]
The color structure of $Q\bar{Q}$ and $q\bar{q}$ should be octet
obviously. However, the color (re)coupling is irrelevant to our
analysis here and will not be explicitly presented. Furthermore, we
will assume no spatial excitation of any quark. Hence all quarks have
zero orbital angular momentum. Then the quantum numbers $J^{PC}$ can be
determined easily for $J_1=0,1$. We have following four $V^{\soge}$
interaction channels:
\[
J_1=0: 1^{+-}; \quad J_1=1: 0^{++}, 1^{++}, 2^{++}.
\]
Finally, we can recover the molecular states by the angular momentum
recoupling. We obtain
\begin{itemize}
\item $0^{++}$
  \[
  [(Q\bar{Q})_1(q\bar{q})_1]_0 = -\frac{\sqrt3}2 (Q\bar{q})_0(q\bar{Q})_0
  -\frac12 [(Q\bar{q})_1(q\bar{Q})_1]_0,
  \]
\item $1^{++}$
  \[
  [(Q\bar{Q})_1(q\bar{q})_1]_1 =-\frac1{\sqrt2} (Q\bar{q})_1(q\bar{Q})_0
  + \frac1{\sqrt2} (Q\bar{q})_0(q\bar{Q})_1,
  \]
\item $1^{+-}$
  \[
  (Q\bar{Q})_0(q\bar{q})_1 = \frac12 (Q\bar{q})_1(q\bar{Q})_0
  + \frac12 (Q\bar{q})_0(q\bar{Q})_1 
  + \frac1{\sqrt2} [(Q\bar{q})_1(q\bar{Q})_1]_1,
  \]
\item $2^{++}$
  \[
  [(Q\bar{Q})_1(q\bar{q})_1]_2 = [(Q\bar{q})_1(q\bar{Q})_1]_2
  \].
\end{itemize}
Then the factor of interaction strength can be read from the
coefficients, which we present in table~\ref{table1}. We see that the
$V^{\soge}$ interaction favors to form the molecular states with
$J^{PC}=1^{++},2^{++}$.
\begin{table}
  \caption{\label{table1}%
    The list of spin factor of $V^{\soge}$ interaction strength in a molecular 
    state $J^{PC}$ made of two meson
    states $J_1^{P_1}$ and $J_2^{P_2}$. $c.c.=\text{charge conjugation}$.}
  \begin{ruledtabular}
    \begin{tabular}{cc|c}
      $J_1^{P_1}J_2^{P_2}$ & $J^{PC}$ & Spin factor \\\hline
      $0^{-}0^{-}$ & $0^{++}$ & $\frac34$ \\
      $1^{-}1^{-}$ & $0^{++}$ & $\frac14$ \\
      $1^{-}0^{-} + c.c.$ & $1^{++}$ & $1$ \\
      $1^{-}0^{-} - c.c.$ & $1^{+-}$ & $\frac12$ \\
      $1^{-}1^{-}$ & $1^{+-}$ & $\frac12$ \\
      $1^{-}1^{-}$ & $2^{++}$ & $1$ \\
    \end{tabular}
  \end{ruledtabular}
\end{table}

\section{Numerical Calculation}
\label{sect-3}

To calculate the molecular states, we use the Rayleigh-Ritz variation
principle. The test wave function will be taken to be a series of Gaussian
basis functions. The Gaussian basis functions are often utilized
in variational calculations of atomic and molecular problems. Recently
the method has been also used in the few body system in nuclear and
particle physics \cite{Kameyama:1989zz, Varga:1996zz, Brink:1998as}.

In our case of the $Q\bar{q}q\bar{Q}$ molecular state, the test wave
function of a molecular state between two clusters of $q\bar{q}$ meson
states is a series
\begin{equation}
  \label{eq-var-4}
  \psi_{1234}(r_{12},r_{34},r_{1234}) = \sum_{i} 
  \alpha_{1234}^i \psi_{12}(r_{12})\psi_{34}(r_{34})
  \exp(-\beta_{1234}^i r_{1234}^2),
\end{equation}
where $\bm{r}_1$, $\bm{r}_2$, $\bm{r}_3$ and $\bm{r}_4$ are the
coordinates of $Q$, $\bar{q}$, $q$ and $\bar{Q}$,
respectively. $\bm{r}_{ij} = \bm{r}_i -\bm{r}_j$.  $r_{1234}$ is the
distance between the two meson clusters
\begin{equation}
\bm{r}_{1234} = \frac{m_Q\bm{r}_1+m_q\bm{r}_2}{m_Q+m_q}
-\frac{m_q\bm{r}_3+m_Q\bm{r}_4}{m_q+m_Q}. 
\end{equation}
$\psi_{ij}(r_{ij})$ is the meson wave function which is also taken to
be a Gaussian function series
\begin{equation}
  \label{eq-var-2}
  \psi_{ij}(r_{ij}) = \sum_k \alpha_{ij}^k 
  \exp(-\beta_{ij}^k r_{ij}^2) .
\end{equation}

The wave function of a molecular state is determined by the variation
principle in two steps. We first determine the wave function
(\ref{eq-var-2}) of each meson cluster.  Then the meson cluster
functions $\psi_{ij}$ are fixed in (\ref{eq-var-4}) to obtain the wave
function of the molecular state and their masses.

To reduce the amount of computation, the parameters $\beta^i$ and
$\alpha^i$ in a Gaussian function series are determined in two steps by
one-dimensional minimization.  We first determined a average $\beta$
value using a single Guassian function. Then a set $\{\beta^i\}$ of
$2N+1$ elements is generated from scaling the $\beta$ value up and
down by a scale factor $s$ \cite{Brink:1998as}:
\begin{equation}
  \beta^i = \beta s^{i-N}
\end{equation}
where $i=0,1,...,2N$. The coefficients $\alpha^i$ are determined by
diagonalization the model Hamiltonian in the $2N+1$-dimensional space
spanned by these $2N+1$ different Gaussian functions. The final values
of $\beta^i$ and $\alpha^i$ and the mass of tetra quark states are
determined by searching the scale factor $s$ for a minimum of system
energy.

In this way, we have calculated the possible $0^{++}$(the combination
of two $0^-$ mesons with the spin factor $\frac34$), $1^{++}$ and
$2^{++}$ molecular states in table~\ref{table1}. We have calculated both
the charmonium-like states $c\bar{q}q\bar{c}$ which will be compared to
recent experimental results and the bottomium-like states
$b\bar{q}q\bar{b}$ which can be investigated in further experiments.

The results of relevant heavy quark $Q\bar{q}$ mesons are listed in
table~\ref{table-2}. We see that the mass values calculated from the
Bhaduri quark model deviate at most $30$ MeV away from the
experimental data. The difference between the calculation from the
variation method with Gussian function series (cal.~II) and the exact
numerical calculation (cal.~I) is less than $0.5$ MeV. So the
variation method is an impresssive good approximation for the
numerical calculation of the $Q\bar{q}$ conventional mesons.
\begin{table}
  \caption{\label{table-2}%
    Mass of $Q\bar{q}$ mesons. The experimental values are taken from 
    PDG\cite{Amsler:2008zzb}. In calculation I, the mass is obtained by
    solving the Schr\"odinger equation. In calcalation II, the mass is obtained
    by the variation method with Gaussian test functions using $N=3$.
  }
  \begin{ruledtabular}
    \begin{tabular}{c|ddd}
      meson state & 
      \multicolumn{1}{c} exp. (MeV) & 
      \multicolumn{1}{c} cal. I (MeV) & 
      \multicolumn{1}{c} cal. II (MeV)\\\hline
      $D^{\pm}$ & 1869.62 & 1885.56 & 1886.14 \\
      $D^0$ & 1864.84 & &\\
      $D^*(2007)^0$ & 2006.97 & 2019.96 & 2020.07 \\
      $D^*(2010)^\pm$ & 2010.27 & &\\\hline
      $D_s^\pm$ & 1968.49 & 1995.78 & 1996.36 \\
      $D_s^{*\pm}$ & 2112.3 & 2101.2 & 2101.4 \\\hline
      $B^\pm$ & 5279.15 & 5300.84 & 5301.18\\
      $B^0$ & 5279.53 & &\\
      $B^*$ & 5325.1 & 5350.3 & 5350.5 \\\hline
      $B_s^0$ & 5366.3 & 5371.9 & 5372.4 \\
      $B_s^*$ & 5412.8 & 5413.3 & 5413.6
    \end{tabular}
  \end{ruledtabular}
\end{table}

The results of the molecular states are given in table~\ref{table-3}
and table~\ref{table-4}.  Here the molecular state is presented
by its binding energy:
\begin{equation}
  E_b = M_1 + M_2 - M_X,
\end{equation}
and the rms radius $\langle r^2 \rangle^{1/2}$, where $M_1$, $M_2$ are
masses of the two compound mesons, $M_X$ is the mass of the molecular
state.

\begin{table}
  \caption{\label{table-3}%
    Binding energy $E_b$ (in MeV) of molecular states vs paramter $f$.
  }
  \begin{ruledtabular}
    \begin{tabular}{c|ddddd}
      $f$ & -0.8 & -1.0 & -1.5 & -2.0 & -3.0 \\\hline
      $(DD)0^{++}$ & - & - & 11.8 & 28.1 & 71.7 \\
      $(DD^*)1^{++}$ & 1.8 & 6.2 & 24.2 & 48.7 & 109.3 \\
      $(D^*D^*)2^{++}$ & - & 5.4 & 21.4 & 43.5 & 98.1 \\\hline
      $(D_sD_s)0^{++}$ & - & - & - & - & 5.3 \\
      $(D_sD^*_s)1^{++}$ & - & - & - & - & 15.4 \\
      $(D^*_sD^*_s)2^{++}$ & - & - & - & - & 14.0 \\\hline
      $(BB)0^{++}$ & 8.8 & 15.3 & 35.2 & 58.5 & 110.9 \\
      $(BB^*)1^{++}$ & 16.8 & 26.8 & 55.8 & 88.5 & 160.4 \\
      $(B^*B^*)2^{++}$ & 16.0 & 25.5 & 53.4 & 84.7 & 153.6 \\\hline
      $(B_sB_s)0^{++}$ & - & - & 3.4 & 9.3 & 25.7 \\
      $(B_sB^*_s)1^{++}$ & - & 1.9 & 8.9 & 18.9 & 44.4 \\
      $(B^*_sB^*_s)2^{++}$ & - & 1.8 & 8.5 & 18.2 & 42.9
    \end{tabular}
  \end{ruledtabular}
\end{table}

\begin{table}
  \caption{\label{table-4}%
    rms $\langle r^2 \rangle^{1/2}$ (in fm) of 
    molecular states vs paramter $f$.
  }
  \begin{ruledtabular}
    \begin{tabular}{c|ddddd}
      $f$ & -0.8 & -1.0 & -1.5 & -2.0 & -3.0 \\\hline
      $(DD)0^{++}$ & - & - & 1.33 & 1.00 & 0.76 \\
      $(DD^*)1^{++}$ & 2.67 & 1.68 & 1.07 & 0.86 & 0.69 \\
      $(D^*D^*)2^{++}$ & - & 1.79 & 1.12 & 0.91 & 0.72 \\\hline
      $(D_sD_s)0^{++}$ & - & - & - & - & 1.63 \\
      $(D_sD^*_s)1^{++}$ & - & - & - & - & 1.12 \\
      $(D^*_sD^*_s)2^{++}$ & - & - & - & - & 1.17 \\\hline
      $(BB)0^{++}$ & 1.11 & 0.94 & 0.74 & 0.64 & 0.54 \\
      $(BB^*)1^{++}$ & 0.92 & 0.80 & 0.66 & 0.58 & 0.50 \\
      $(B^*B^*)2^{++}$ & 0.94 & 0.82 & 0.67 & 0.59 & 0.51 \\\hline
      $(B_sB_s)0^{++}$ & - & - & 1.42 & 1.01 & 0.74 \\
      $(B_sB^*_s)1^{++}$ & - & 1.77 & 1.03 & 0.81 & 0.63 \\
      $(B^*_sB^*_s)2^{++}$ & - & 1.81 & 1.05 & 0.83 & 0.64
    \end{tabular}
  \end{ruledtabular}
\end{table}

We can see from tables~\ref{table-3} and \ref{table-4} that indeed the
$1^{++}$ and $2^{++}$ molecular states are bound preferable since the
$V^{\soge}$ are strong in these two channels. The binding of $1^{++}$
states are slightly deeper and tighter than that of $2^{++}$'s. So from our
model calculation, the $1^{++}$ molecular states should be easily
observed in experiments.

First, let us look at the $D^{(*)}D^{(*)}$ sector. If the $X(3872)$ is
a molecular state of $D$ and $D^*$, the binding energy can be estimated
from experimental data\cite{Amsler:2008zzb}
\[
E_b = M_D+M_{D^*} - M_X = 0.3 \text{ MeV},
\]
which is very small. Accounting the uncertainty of approximation in our
model calculation, we think the reasonable range of $f$ value should
be around $-0.8\sim -1.0$.  If it is small $f\sim -0.8$, then it may
be the only molecular states in the $D^{(*)}D^{(*)}$ sector. If $f\sim
-1.0$, the $2^{++}$ state may also exists. From our numerical
calculation, it mass should be about
\[
M=M_{D^*}+M_{D^*}-E_b \approx 4008.5 \text{MeV}.
\]
The Belle collaboration has reported a $X(3930)$ state of $2^{++}$ at
mass $M=3929$ MeV\cite{Uehara:2005qd}, which is a candidate of the
$c\bar{c}$ charmonium $2^3P_2$ excited state $\chi_{c2}'$. However its
mass is aboult $40$ MeV below the quark model calculations (in the
Bhaduri model, the $\chi_{c2}'$ mass is $M=3963.53$MeV). Several
authors have discussed the mass shifts of the coupled channel effect
\cite{Kalashnikova:2005ui,Pennington:2007xr,Li:2009ad,Ortega:2009hj}.  
If the above $2^{++}$ molecular state exists, the final state
interaction will be important.

Now, we turn to the possible molecular states in other sectors.  We
observe that in $D^{(*)}_SD^{(*)}_S$ sector, there are no such
molecular states due to the mass increase of $m_s$ of the light quark
pair. On the other hand, in the $B^{(*)}B^{(*)}$ sector, the binding
energy increased as the mass increase of $m_b$ of the heavy quark
pair. So these bottomium-like molecular states with $1^{++}$,
$2^{++}$, and even $0^{++}$ may also exists. We can also observe the
$B^{(*)}_SB^{(*)}_S$ molecular states $1^{++}$ and $2^{++}$ if $f\sim
-1.0$.

\section{Summary}

The quark model is extended by introducing the $s$-channel one gluon
exchange interaction. The interaction has no effect on the inner quark
structure of conventional $q\bar{q}$ mesons and $qqq$ baryons. Since
the interaction is short ranged --- a smeared $\delta$-interaction,
the effect on the long ranged hadron-hadron interaction is expected to be very
small. So the significant effect of this interaction is only in the so
called hidden flavor states of multi-quark system. 

We have calculated the heavy quark molecular states of
$qQ\bar{Q}\bar{q}$ with $Q=c,b$ and $q=u,d,s$. We find that the
interaction can be strong enough to bind the $1^{++}$ and $2^{++}$
states, and possibly $0^{++}$ states. To compare with the recent
experiments, the $X(3872)$ is a candidate of the $DD^{*}$ molecular
state and the $X(3930$ may be the $\chi_{c2}'$ state which couples to
the $D^*D^*$ $2^{++}$ molecular state. The calculation shows that it
is more easy to bind the bottomium-like molecular states. Thus we
expect that the similiar bottomium-like molecular states of $1^{++}$
and $2^{++}$ should also exist.


\begin{acknowledgments}
  We would like to thank Shi-Lin Zhu for
  useful discussions. 
  This work was supported by the National Natural Science Foundation
  of China under Grants 10675008.
\end{acknowledgments}


\end{document}